\pdfoutput=1
\documentclass[ALICE,manyauthors]{cernphprep}
\usepackage[comma,square,numbers,sort&compress]{natbib}
\usepackage{hyperref}
\usepackage{lineno}
\usepackage{xcolor}
\usepackage{floatrow}
\newcommand{\pp}{\ensuremath{\mathrm {p\kern-0.05em p}}}
\newcommand{\PbPb}{\ensuremath{\mbox{Pb--Pb}}}
\newcommand{\pPb}{\ensuremath{\mbox{p--Pb}}}

\newcommand{\sqrtS}{\ensuremath{\sqrt{s}}~}
\newcommand{\sqrtSnn}{\ensuremath{\sqrt{s_{\mathrm{NN}}}}~}
\newcommand{\sqrtSE}[2][TeV]{$\sqrtS = #2\,\mathrm{~#1}$}
\newcommand{\sqrtSnnE}[2][TeV]{$\sqrtSnn = #2\,\mathrm{~#1}$}
\newcommand{\pt}{\ensuremath{p_{\mathrm{T}}}}
\newcommand{\meanpt}{\ensuremath{\langle p_{\mathrm{T}}\rangle}}

\newcommand{\dndy}{\ensuremath{\mathrm{d}N/\mathrm{d}y}}

\newcommand{\avgdndeta}{\ensuremath{\langle\mathrm{d}N_{\rm ch}/\mathrm{d}\eta\rangle}}

\newcommand{\acceff}{\ensuremath{\mathrm{Acc} \times \epsilon}}
\newcommand{\btwo}{\ensuremath{B_{2}}}
\newcommand{\ptOa}{$p_{\rm T}/A$}

\newcommand{\dbar}{\ensuremath{\rm\overline{d}}}

\newcommand{\MeVc}{\ensuremath{\mathrm{MeV}\kern-0.05em/\kern-0.02em c}}
\newcommand{\MeVcSq}{\ensuremath{\mathrm{MeV}\kern-0.05em/\kern-0.02em c^2}}
\newcommand{\GeVc}{\ensuremath{\mathrm{GeV}\kern-0.05em/\kern-0.02em c}}
\newcommand{\GeVcSq}{\ensuremath{\mathrm{GeV}\kern-0.05em/\kern-0.02em c^2}}

\begin{document}%

\begin{titlepage}
\PHyear{2019}
\PHnumber{024}      
\PHdate{15 February}  
%

\title{Multiplicity dependence of (anti-)deuteron production in pp collisions at $\sqrt{s}$  = 7 TeV}
\ShortTitle{Multiplicity dependence of (anti-)deuteron in pp collisions}   

\Collaboration{ALICE Collaboration\thanks{See Appendix~\ref{app:collab} for the list of collaboration members}}
\ShortAuthor{ALICE Collaboration} 

\begin{abstract}
In this letter, the production of deuterons and anti-deuterons in pp collisions at $\sqrt{s} = 7$ TeV is studied as a function of the charged-particle multiplicity density at mid-rapidity with the ALICE detector at the LHC. 
Production yields are measured at mid-rapidity in five multiplicity classes and as a function of the deuteron transverse momentum (\pt). The measurements are discussed in the context of hadron--coalescence models. The coalescence parameter \btwo, extracted from the measured spectra of (anti-)deuterons and primary (anti-)protons, exhibits no significant \pt-dependence for $\pt < 3$~\GeVc, in agreement with the expectations of a simple coalescence picture. At fixed transverse momentum per nucleon, the \btwo~parameter is found to decrease smoothly from low multiplicity \pp~to \PbPb~collisions, in qualitative agreement with more elaborate coalescence models. The measured mean transverse momentum of (anti-)deuterons in pp is not reproduced by the Blast-Wave model calculations that simultaneously describe pion, kaon and proton spectra, in contrast to central \PbPb~collisions. The ratio between the \pt-integrated yield of deuterons to protons, d/p, is found to increase with the charged-particle multiplicity, as observed in inelastic pp collisions at different centre-of-mass energies. The d/p ratios are reported in a wide range, from the lowest to the highest multiplicity values measured in pp collisions at the LHC.

\end{abstract}
\end{titlepage}
\setcounter{page}{2}


\section{Introduction}
The production of light nuclei and anti-nuclei in elementary collisions has been described by phenomenological models in which nucleons coalesce into nuclei \cite{Butler:1963, Kapusta:1980, Csernai:1986, Scheibl:1998}. 
According to these models, a pair of independent final-state nucleons that are nearby in space and have similar velocities can transfer energy to the rest of the system to form a deuteron or an anti-deuteron.
The production rate of the (anti-)deuteron obtained by coalescence is thus related to those of its constituent protons and neutrons. In order to provide a quantitative description of this process the coalescence parameter \btwo, which relates the deuteron production to the square product of nucleon yields, is extracted. 
These models have successfully been tested with deuteron and anti-deuteron production measured in pp collisions at the CERN ISR \cite{Alper:1973my, Henning:1977mt} and Tevatron \cite{Alexopoulos:2000jk}, photo-production and deep inelastic scattering of electrons at HERA \cite{Aktas:2004pq, Chekanov:2007mv}, electron-positron collisions at ARGUS \cite{Argus}, BaBar \cite{Babar}, CLEO \cite{Asner:2006pw} and at LEP \cite{Schael:2006fd}. 
Results on the production of light (anti-)nuclei in inelastic pp collisions at \sqrtSE{0.9, 2.76~${\rm and}$~7}
have been reported by the ALICE Collaboration in~\cite{ALICE:deuteronppPbPb2015, ALICE:nucleipp2017} and the validity of coalescence models~\cite{Butler:1963, Kapusta:1980, Csernai:1986, Scheibl:1998} at the Large Hadron Collider (LHC) has also been discussed.
Light nuclei and their anti-matter counterparts are rarely produced in elementary reactions. In pp collisions at LHC energies, 
the cost to add one constituent nucleon to a nucleus amounts to a reduction factor of the yield (also called ``penalty factor'') of about 1000 \cite{ALICE:nucleipp2017}. 
Heavy-ion collisions, on the other hand, constitute a more abundant source of light (anti-)nuclei, as reported by ALICE \cite{ALICE:deuteronppPbPb2015, ALICE:deuteronflow2017, ALICE:alphaPbPb2017}. A penalty factor of about 300 has been extracted in central \PbPb~collisions at the LHC \cite{ALICE:alphaPbPb2017}.

In \PbPb~collisions, the yields of light (anti-)nuclei up to the mass number $A$ = 4 have been successfully described together with other light-flavour hadrons in the thermal-statistical approach with one common chemical freeze-out temperature \cite{Andronic:nuclei2011, Andronic:2017, ALICE:alphaPbPb2017}. 
Compared to hydrodynamic-inspired models (e.g. Blast-Wave model~\cite{BW:1993}), the measured deuteron \pt~spectra and elliptic-flow coefficient ($v_{2}$) suggest common kinetic freeze-out conditions for deuterons and primary pions, kaons and protons \cite{ALICE:deuteronppPbPb2015, ALICE:deuteronflow2017}. 
Furthermore, the relative deuteron-to-proton yields (d$/$p) increase by about a factor two from inelastic pp to central Pb--Pb collisions,
where the values \cite{ALICE:deuteronppPbPb2015} are in agreement with the statistical-thermal model \cite{Andronic:2017}. 
A coalescence approach that neglects the size of the particle emitting source (hereafter denoted as ``simple coalescence'') fails in reproducing the deuteron \btwo~and $v_{2}$ measured in \PbPb~collisions \cite{ALICE:deuteronppPbPb2015, ALICE:deuteronflow2017}.
A formulation of the coalescence model that takes into account the size of the particle-emitting source has been proposed to describe the behaviour in large systems \cite{Scheibl:1998}.
In smaller systems one has to consider that the size of the deuteron may be as large as or even larger than the size of the emitting source. 

The abundances of nuclei are very sensitive to the freeze-out conditions, to the dynamics, and the size of the emitting source. For these reasons, a systematic comparison of the production of light nuclei across different collision systems and, in particular, in events with similar final-state multiplicity but very different initial conditions and collision geometry can shed light on the production mechanisms. 
Thanks to the high statistics data sample collected by ALICE, the deuteron and anti-deuteron production in pp collisions can be studied differentially as a function of the charged-particle multiplicity and the transverse momentum (\pt), complementing the previous measurements in pp and Pb--Pb collisions.

This letter is organised as follows: in Sec.~2 the experimental apparatus, the analysis technique and the estimation of the systematic uncertainties are described. The results on multiplicity dependent \pt-differential and \pt-integrated yields and the anti-deuteron over deuteron ratio are reported in Sec.~3, which also contains a detailed discussion of the results. Conclusions follow in Sec.~4.

\section{Experimental details}

\subsection{The ALICE detector}
\label{sec:detector}
A comprehensive description of the ALICE apparatus and its performance can be found in \cite{Aamodt:2008zz, Abelev:2014ffa}. 
In this section, the detectors used for the analysis discussed in this paper are described. 
Deuteron spectra are measured at mid-rapidity ($|y|<0.5$) relying on the
tracking and particle identification (PID) capabilities of the central-barrel detectors, which are located in a solenoid magnet providing a $B$ = 0.5 T field, parallel to the beam direction ($z$-axis in the ALICE reference frame).

From the innermost radius of 3.9 cm (distance from the centre of the beam vacuum pipe) to the outermost radius of 43 cm,
the Inner Tracking System (ITS) includes two layers of Silicon Pixel Detector (SPD),
two Silicon Drift Detector (SDD) layers, and two Silicon Strip Detector (SSD) layers.
The different ITS sub-systems have full azimuth and a common pseudorapidity coverage of $|\eta| < 0.9$ in the acceptance. 
The spatial precision of the ITS, its proximity to the beam pipe, and its very low material budget \cite{Aamodt:2010aa} enable a precise determination of the primary vertex and of the track impact parameter (i.e. the distance of closest approach of the track to the primary vertex) in the transverse plane, for which a resolution better than 75~$\mu$m  is achieved for tracks with \pt~\textgreater~1~\GeVc~\cite{Aamodt:2010aa}. 

The Time Projection Chamber (TPC) is the main tracking device of the experiment and surrounds the ITS with an active volume ranging from 85 cm to 247 cm in radius with full azimuthal coverage in the pseudorapidity interval $|\eta| < 0.9$. It provides up to 159 space points to determine the particle trajectory and measure its momentum. Moreover, the specific ionisation energy-loss of particles inside the TPC volume is measured with a resolution of 5$\%$ in pp collisions, exploited here for PID.

The Time-Of-Flight (TOF) system~\cite{TOFperformance2013}, an array of 1593 Multi-gap Resistive Plate Chambers, completes the set of detectors used for PID in the analysis presented in this letter. It is located at a radial distance of about 3.8 m, covering full azimuth in the pseudorapidity interval $|\eta| < 0.9$.
The event time of the collision is obtained on an event-by-event basis either using the TOF detector, or the T0 detector, or a combination of the two \cite{Adam:2016ilk}. The T0 detector consists of two arrays of Cherenkov counters, located on both sides of the interaction point at $z = 350$ cm and $z = -70$ cm from the nominal vertex position.
The time-of-flight of the particles is determined with a resolution of about 120 ps in pp collisions.

Between the TOF and the TPC, the Transition Radiation Detector (TRD) is positioned at a radial distance between 2.9 and 3.7 m from the beam axis, with pseudorapidity coverage of $|\eta| < 0.8$.
Since 2014, all eighteen TRD supermodules are installed, covering full azimuth. In 2010, when the data used for the analysis presented here were collected, only seven sectors were present.
Although the TRD is not used in this analysis, its detector material plays a role in the efficiency corrections, described in Sec.~\ref{sec:efficiency}.


The V0 detector consists of two scintillator arrays built around the beam pipe on either side of the interaction point at $z=329$ cm and $z=-88$ cm, and covering the pseudorapidity ranges $2.8 \le \eta \le 5.1$ (V0-A) and -3.7 $\le \eta \le$ -1.7 (V0-C). This detector is used for triggering and background suppression. It is also employed for classifying events according to multiplicity, as further detailed in the next section.

\subsection{Event selection and multiplicity classes}\label{sec:evSel}

The analysis is based on a data sample of 237 million minimum-bias triggered pp collisions at \sqrtSE{7}. The minimum-bias trigger requires a hit in either the V0 or the SPD, in coincidence with the crossing of proton bunches from the two beams. The timing information provided by the V0 detector as well as the correlation between the SPD hit multiplicity and the number of SPD track segments pointing to the primary vertex are used offline to reject the contamination from beam-gas events, achieving a purity of the minimum-bias event sample of 99.7$\%$ as estimated in \cite{Abelev:2014ffa}.
The pileup rejection is performed by rejecting offline the events with more than one reconstructed vertex in the SPD.  
The residual fraction of events with pileup ranges from about $10^{-4}$ to $10^{-2}$ for the lowest and highest multiplicity classes, respectively.
Events are also required to have a primary vertex reconstructed by the SPD within $\pm$ 10 cm from the nominal interaction point along the beam direction. The sample selected with the above criteria contains 172 million events.

The results are reported for an event class (INEL\textgreater0) characterised by at least one charged particle being produced in the pseudorapidity interval $\vert \eta \vert$ \textless~1, corresponding to about 75$\%$ of the total inelastic cross-section. 
INEL\textgreater0 events are selected experimentally by requiring that at least one track segment (tracklet) is reconstructed in the SPD.
This selection can be affected by inefficiencies associated with the tracklet reconstruction. Thus the selected number of events used for the normalisation of the yields is corrected for the 8.5$\%$ loss due to inefficiency in the lowest multiplicity class and for less than 1.2$\%$ loss for all other classes, as estimated in \cite{ALICE:npstrange2017}.

In order to study deuteron production as a function of  multiplicity, the selected events are classified using the ``V0M'' forward multiplicity estimator, based on the total energy deposited in both the V0 scintillator arrays (V0-A and V0-C). The V0M amplitude is linearly proportional to the total number of charged particles produced in the V0 detectors acceptance. Since deuteron production is measured at mid-rapidity, an independent estimator is preferred as an event classifier to avoid auto-correlation biases.
In each V0M event class the average charged-particle multiplicity density (\avgdndeta) is measured at mid-rapidity and results are reported in the following as a function of \avgdndeta.

For the event classes relevant for this analysis, the values of \avgdndeta~and the fraction of the INEL\textgreater0 cross section are reported in Tab.~\ref{tab:multiplicity}. Roman numerals are used to indicate each of the ten event classes in which the measurement of other light-flavour hadron yields, and protons in particular, have been performed as reported in \cite{ALICE:npstrange2017, ALICE:longpaperpp2018}.
Considering the deuteron statistics needed for the present analysis, some of these classes have been combined as indicated in the table.  
 
\begin{table}[!t]
\begin{center}
\begin{tabular}{ccc}
\hline
\hline
Multiplicity Class & $\sigma$/$\sigma_{\rm INEL > 0}$ & \avgdndeta \\
\hline
I+II       & 0 - 4.7 $\%$ & $17.47 \pm 0.52$\\
III      & 4.7 - 9.5 $\%$ & $13.50 \pm 0.40$\\
IV+V      & 9.5 - 19 $\%$ & $10.76 \pm 0.30$\\
VI+VII     & 19 - 38 $\%$  & $7.54 \pm 0.23$\\
VIII+IX+X & 38 - 100 $\%$ & $3.30 \pm 0.13$\\
I to X     & 0 - 100 $\%$ & $5.96 \pm 0.23$\\
\hline
\hline
\end{tabular}
\medskip 
\caption{Charged-particle multiplicity (\avgdndeta) measured at mid-rapidity ($\left|\eta\right|$ \textless~0.5) and its corresponding fraction of the INEL\textgreater0 cross section ($\sigma$/$\sigma_{\rm INEL>0}$) for each of the multiplicity classes selected with the V0M estimator and relevant for this analysis, indicated by roman numerals \cite{ALICE:npstrange2017}. The uncertainties are the square-root of the sum in quadrature of statistical and systematic contributions and represent one standard deviation.}
\label{tab:multiplicity}
\end{center}
\end{table}

\subsection{Track selection and particle identification}
\label{sec:PID}
In order to ensure good quality, tracks are selected according to the following criteria.
For each track, at least two reconstructed points are required in the ITS (including at least one in the SPD)
and 70 out of a maximum of 159 in the TPC.
The track-fit quality is assured by requiring the $\chi^{2}$ per space point in the TPC to be less than 4. 
Daughter tracks from reconstructed kinks in the TPC volume are rejected in order to keep only tracks pointing to the primary vertex.
To limit the contamination from secondary particles from material (see Sec.~\ref{sec:secondaries}),
requirements are imposed on the Distance of Closest Approach of each track to the primary vertex along the beam direction
(DCA$_z$) and in the transverse plane (DCA$_{xy}$) to be less than 1~cm and 0.1~cm, respectively.
The fiducial pseudo-rapidity region is defined as $|\eta| < 0.8$, which ensures a uniform acceptance in the detectors involved. 

\begin{figure}[!h]
\begin{center}
\includegraphics[scale=0.5]{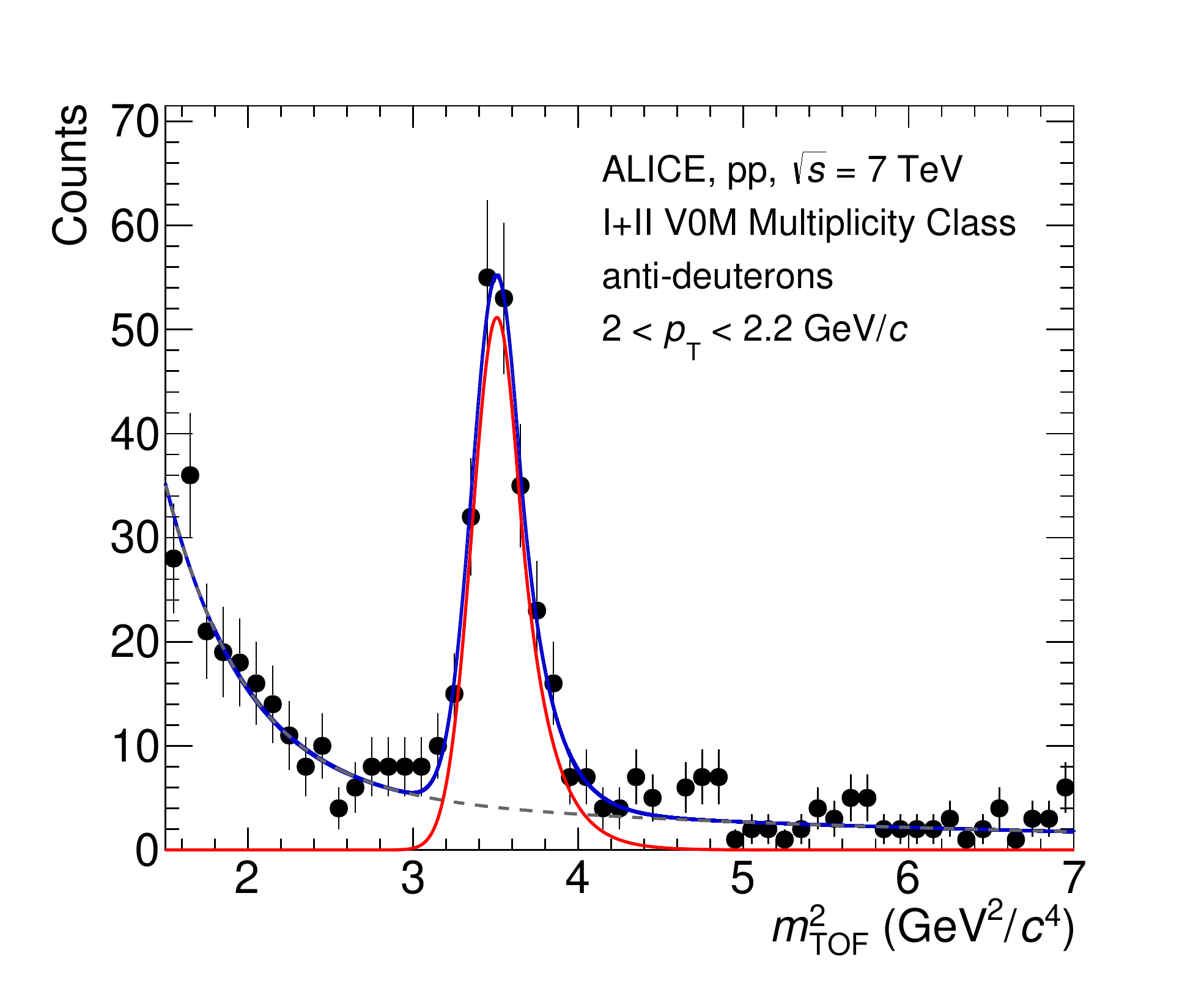}
\caption{TOF squared-mass distribution ($m_{\rm TOF}^2$) around the anti-deuteron peak for a selected \pt~interval and in the highest multiplicity class.
The solid red line represents a fit of a Gaussian function plus an exponential right tail to the anti-deuteron signal, the grey dashed line the fit of the background performed using the sum of two exponential functions, and the solid blue line is the sum of the signal and background components.} 
\label{fig:dbarTOF}
\end{center}
\end{figure}

The identification of (anti-)deuterons is achieved by exploiting the measurement of their specific ionisation energy-loss, provided by the TPC, and via the measurement of the time-of-flight of the particles, performed with the TOF.
Due to the different acceptance of the two detectors, the TPC is used without the TOF for \pt~\textless~1~\GeVc, where the separation of deuterons from light hadrons is very effective.
Deuterons and anti-deuterons are selected by requiring an energy loss compatible, within $\pm3\sigma$, with the value expected for particles having the mass and charge of the deuteron, where $\sigma$ is the resolution of the particle energy loss in the TPC.
For $\pt > 1$ \GeVc, TOF information is required together with that from the TPC.
The squared mass of the particles, $m_{\rm TOF}^2 = p^2 \, (t_{\rm TOF}^2/L^2 - 1/c^2)$, is then determined
from the measured time-of-flight ($t_{\rm TOF}$), the momentum ($p$) and the track length ($L$), after the $3\sigma$ selection on the particle energy-loss in the TPC. 
Figure~\ref{fig:dbarTOF} shows an example of the obtained $m_{\rm TOF}^2$ distribution around the anti-deuteron peak for a selected \pt~interval and in the highest multiplicity class (I+II). 
The $m_{\rm TOF}^2$ distribution is fitted using a Gaussian function with an exponential tail towards higher masses for the signal that reflects the TOF detector time response \cite{TOFperformance2013}. To describe the background the sum of two exponential functions is used. They account for those tracks erroneously associated to a TOF hit and for the tail of the (anti-)proton signal.
For both the TPC-only and TOF-TPC analyses the yields of deuterons and anti-deuterons are separately extracted in each \pt~interval and for each multiplicity class. 

\subsection{Rejection of secondary deuterons}
\label{sec:secondaries}
The sample of identified deuterons is contaminated by those that originate from interactions of primary particles with the detector material, e.g. knock-out or pick-up, which are highly suppressed for anti-deuterons.
The corresponding correction, only for matter, is estimated as in \cite{ALICE:nucleipp2017} and is based on a fit to the distribution of the DCA$_{xy}$. The latter is determined as the sum of two contributions: the signal of primary deuterons
appears as a Gaussian-like peak centred around zero whereas secondary nuclei contribute to the flat underlying background.
The fraction of secondary deuterons is about 40$\%$ at \pt~$\simeq$~0.6~\GeVc~and decreases exponentially as the transverse momentum increases until it becomes smaller than 5$\%$ above 1.4~\GeVc.   
It is observed that this does not depend on multiplicity and therefore a correction based on the multiplicity-integrated data sample is used to minimise the statistical uncertainties.

\subsection{Acceptance and efficiency} 
\label{sec:efficiency}
After subtracting the contamination from secondary particles, raw yields are corrected for acceptance and tracking efficiency (\acceff).
This correction allows one to account for the limited acceptance of the detectors, the particle absorption in the detector material -- mainly due to energy loss and multiple-scattering processes -- and the partial inefficiencies due to detector dead zones and inactive readout channels.
The \acceff~is computed by using Monte Carlo (MC) generated events. 
Standard event generators for pp collisions, e.g. PHOJET~\cite{PHOJET:1995} or PYTHIA~\cite{PYTHIA:2006} do not consider the production of nuclei. 
To include light (anti-)nuclei, these are injected into underlying PHOJET events with flat momentum and rapidity distributions. 
The ALICE detector description is based on the GEANT3 particle transport code~\cite{Geant3:1994}. 
As discussed in \cite{ALICE:deuteronppPbPb2015}, GEANT3 includes only an approximate description of the interactions of light nuclei with the detector material.
The \acceff~is reduced by 6$\%$ when TOF PID is used, due to the extra \mbox{(anti-)deuterons} lost because of hadronic interactions that GEANT3 does not account for.
This correction is based on the fraction of (anti-)deuterons absorbed in the TRD modules installed between TPC and TOF, studied in data and MC simulations. 
More details can be found in \cite{ALICE:nucleipp2017}.

As already mentioned in Sec. \ref{sec:evSel}, the \pt-differential yields are normalised to INEL\textgreater0 events.
Raw yields need to be further corrected for the amount of (anti-)deuteron signals lost because of the event selection.
This correction is expected to be dependent on multiplicity. Simulations enriched with nuclei, such as those used to determine \acceff, are not appropriate for its estimation, because the mean number of charged particles per event is not well described.  
In this respect, a MC simulation (based on PYTHIA as event generator) that reproduces the charged-particle multiplicity measured in the data can be safely used. Since such simulations do not contain nuclei, the fraction of signal lost in the event selection is estimated for (anti-)deuterons by extrapolating the ones determined for pions, kaons and protons.
This has been done by exploiting the linear dependence of the lost signal as a function of the mass of the particles, which was observed in simulations.
For the lowest multiplicity class, the resulting fraction of deuteron loss is about 4$\%$ at \pt~$\simeq$~0.6~\GeVc~and rapidly decreases as the transverse momentum increases until it becomes smaller than 1$\%$ above 1~\GeVc. For higher multiplicities, the correction is negligible.
    

\subsection{Systematic uncertainties}
There are several contributions to the total systematic uncertainty. Two contributions arise from the particular set of selections applied to the sample of tracks for the analysis and from the particle identification procedure. The rejection of secondary deuterons also introduces an uncertainty. Other significant uncertainties originate from the limited knowledge of the absorption of light (anti-)nuclei in the detector material and of the amount of material itself. The ITS-TPC track matching efficiency is also known with finite precision. 
The normalisation of the \pt-differential yields to INEL\textgreater0 events is an additional source of uncertainty. All contributions to the total systematic uncertainty are summarised in Tab.~\ref{tab:systematics} for the highest multiplicity class (I+II). More details are presented in the following.

The systematic uncertainty related to PID is smaller at low transverse momenta, down to $3\%$ at 0.6 \GeVc, because of a clear separation of the deuteron and anti-deuteron signals in the TPC.
At higher \pt, the presence of the background, which contaminates the signal in the TOF significantly, introduces an additional uncertainty. 
The latter increases gradually from about 3$\%$ at 1 \GeVc~to about 22--$23\%$ for $\pt \approx 3$ \GeVc.
The uncertainty at high transverse momentum, at $\pt \approx 3$ \GeVc, originates mainly from the right tail of the proton squared-mass distribution, which strongly contaminates the (anti-)deuteron signal in the TOF.

In the case of the TPC PID, the systematic uncertainty estimate is based on a variation of the maximum accepted difference between the measured and expected energy-loss value for the (anti-)deuteron-mass hypothesis. In the case of TOF PID, the bin width of the squared-mass distribution and the range of the fit have been varied. At intermediate transverse momenta ($1<p_{\rm T}<1.6~{\rm GeV}/c$), where the background under the (anti-)deuteron signal peak in the $m_{\rm TOF}^2$ distribution is almost negligible, the yield is extracted by bin counting. This result is compared to the one obtained with the fit procedure described in Sec.~\ref{sec:PID} in order to estimate the systematic uncertainty.
The uncertainty resulting from the track selection has been estimated through variations of the specific requirements used in the analysis. 
The rejection of secondary deuterons is also a source of uncertainty at low \pt~while it is negligible for anti-deuterons.
The uncertainty is estimated by varying the maximum $|$DCA$_{z}|$ of the accepted tracks, which has a significant impact on the estimated fraction of primary particles.  
A \pt-independent uncertainty of 3$\%$ is associated with the difference between the ITS-TPC track matching efficiency in data and MC simulations \cite{ALICE:spectraPPmb2015, ALICE:longpaperpp2018}.
The systematic uncertainty related to the normalisation of the spectra to the INEL\textgreater0 event class is found to be not larger than $1\%$  for all multiplicities and transverse momenta. This uncertainty is estimated as the difference between the corresponding proton and deuteron corrections (see Sec.~\ref{sec:efficiency}).

The limited knowledge of the hadronic interaction cross section of the primary particles in the detector material leads to a systematic uncertainty of 6$\%$ uniform in \pt, as estimated in \cite{ALICE:nucleipp2017}.
Moreover, the uncertainty of the material budget contributes with an additional $3\%$ to the total uncertainty. For its evaluation, the effect of varying the relative amount of material by $\pm 10\%$ has been studied through simulations.
All the mentioned contributions have been summed in quadrature. The total systematic uncertainty depends moderately on multiplicity: the relative difference between different multiplicity classes is 20--30$\%$ at most.

\begin{table}[!t]
\begin{center}
\begin{tabular}{lcc}
\hline
\hline
\rule{0pt}{2.5ex}
Source & \multicolumn{2}{c}{d (\dbar)}\\
  \pt     & 0.6~GeV/${\it c}$ & 3~GeV/${\it c}$\\
\hline
\rule{0pt}{2.5ex}
Particle identification & 3$\%$ & 24$\%$ (26$\%$)\\
Track selection & 1$\%$ & 5$\%$\\
Secondary nuclei & 7$\%$ (negl.) & negl.\\
ITS-TPC matching & 3$\%$ & 3$\%$\\ 
Norm. to INEL~\textgreater~0 events & 1$\%$ & negl.\\
Hadronic interactions & 6$\%$ & 6$\%$\\
Material budget & 3$\%$ & 3$\%$\\
\hline
\rule{0pt}{2.5ex}
Total & 11$\%$ (8$\%$) & 26$\%$ (27$\%$) \\
\hline
\hline
\end{tabular}
\medskip 
\caption{Systematic uncertainties on deuteron and anti-deuteron transverse-momentum spectra at low and high \pt~for the highest multiplicity class (I+II). The values in parentheses apply to anti-deuterons and are only given where they differ from those related to deuterons. Otherwise, where it is not explicitly specified, the values are common to particles and anti-particles.}
\label{tab:systematics}
\end{center}
\end{table}

\section{Results and discussion}

\begin{figure}[!t]
\begin{center}
\includegraphics[scale=0.8]{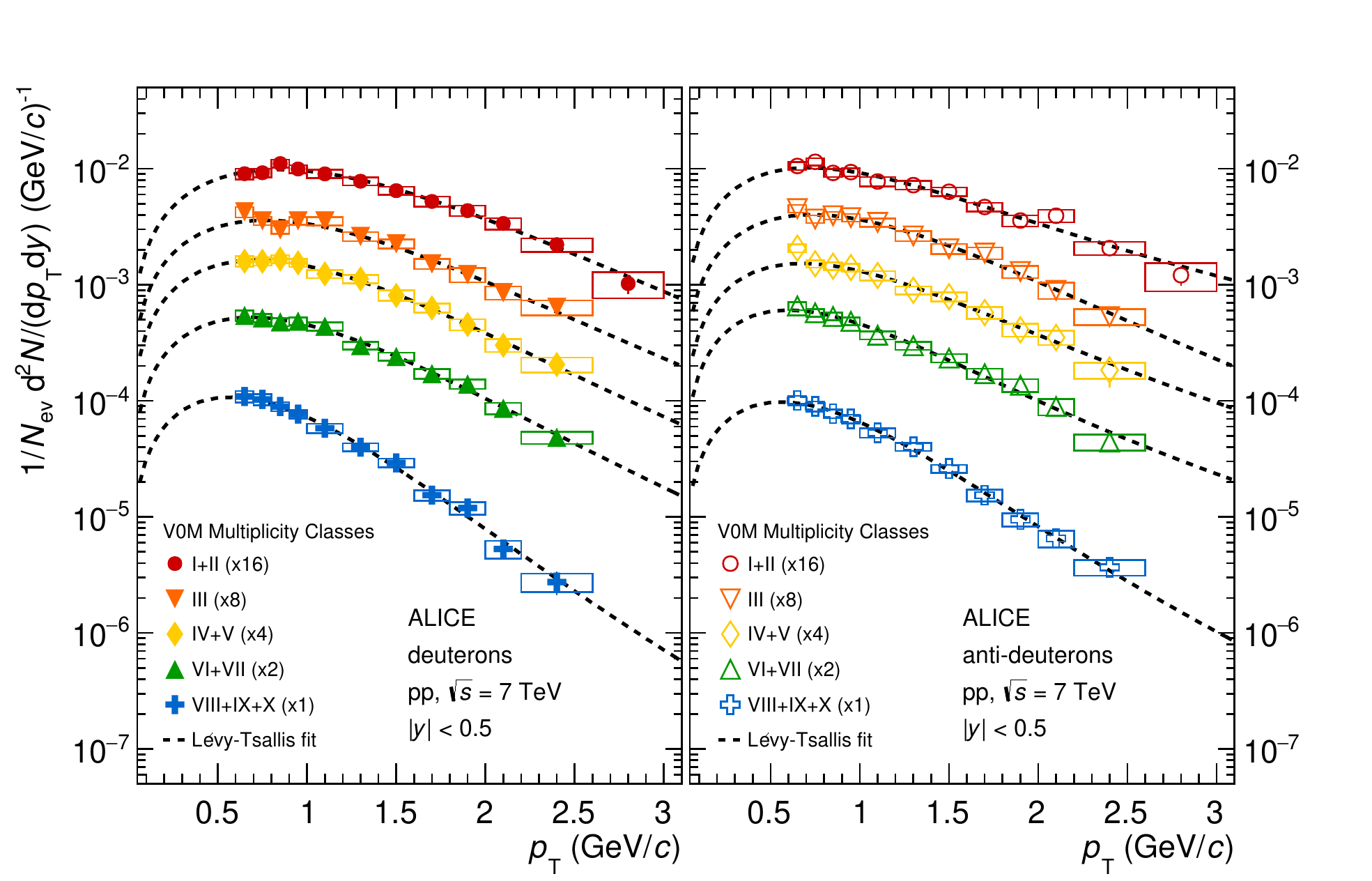}
\caption{Transverse-momentum spectra of deuterons (left) and anti-deuterons (right) measured at mid-rapidity in pp collisions at \sqrtSE{7}~in the considered multiplicity classes. The vertical bars are the statistical uncertainties, the open boxes represent the systematic ones. The dashed lines correspond to individual fits to the data performed with the L\'evy-Tsallis function (see Eq.~\ref{eq:levy}). The spectra have been scaled with the indicated factors for better visibility.} 
\label{fig:spectra_ddbar}
\end{center}
\end{figure}

\begin{table}[!h]
\begin{center}
\begin{tabular}{ccccc}
\hline
\hline
\rule{0pt}{2.5ex}
& Multiplicity Class & ${\rm d}N/{\rm d}y \, (\times 10^{-4})$ & $\mathcal{h}p_{\rm T}\mathcal{i}$ (\GeVc) & Extr. ($\%$)\\
\hline
\rule{0pt}{2.5ex}
  & I+II & $ 10.14 \pm 0.15 \pm 1.17 $ & $1.28 \pm 0.01 \pm 0.06$ & 23\\
  & III & $ 7.01 \pm 0.13 \pm 0.81 $ & $1.19 \pm 0.02 \pm 0.07$ & 27\\
d & IV+V & $ 5.76 \pm 0.08 \pm 0.64 $ & $1.11 \pm 0.01 \pm 0.05$ & 29\\
  & VI+VII & $ 3.55 \pm 0.04 \pm 0.39 $ & $1.05 \pm 0.01 \pm 0.05$ & 30\\
  & VIII+IX+X & $ 1.15 \pm 0.01 \pm 0.17 $ & $0.82 \pm 0.01 \pm 0.05$ & 39\\
\hline
\rule{0pt}{2ex}
                     & I+II & $ 10.87 \pm 0.18 \pm 1.47 $ & $1.47 \pm 0.02 \pm 0.16$ & 31\\
                     & III & $ 7.44 \pm 0.15 \pm 0.82 $ & $1.16 \pm 0.02 \pm 0.06$ & 29\\
${\rm \overline{d}}$ & IV+V & $ 5.68 \pm 0.13 \pm 0.68 $ & $1.17 \pm 0.02 \pm 0.10$ & 31\\
                     & VI+VII & $ 3.88 \pm 0.06 \pm 0.44 $ & $1.05 \pm 0.01 \pm 0.07$ & 35\\
                     & VIII+IX+X & $ 1.07 \pm 0.02 \pm 0.15 $ & $0.85 \pm 0.01 \pm 0.05$ & 39\\
\hline
\hline
\end{tabular}
\medskip 
\caption{\pt-integrated yield, \dndy, and mean transverse momentum, \meanpt, along with the extrapolated fraction (Extr.) of deuterons (top) and anti-deuterons (bottom) 
in pp collisions at \sqrtSE{7}~in different multiplicity classes. The first uncertainty is statistical, the second one is the sum in quadrature of the systematic error and the uncertainty due to the spectrum extrapolation, as described in the text.}
\label{tab:fitSpectraValues}
\end{center}
\end{table}

\subsection{Transverse momentum spectra}
The transverse momentum spectra of deuterons and anti-deuterons in the considered multiplicity classes are shown in Fig.~\ref{fig:spectra_ddbar}, in the left and right panels, respectively.
In order to extrapolate the spectra to low and high \pt, the distributions are individually fitted with the L\'evy-Tsallis function~\cite{Tsallis:1988, ALICE:spectraPPmb2015},

\begin{equation}
\label{eq:levy}
\frac{{\rm d}^{2}N}{{\rm d}p_{\rm T}{\rm d}y} = \frac{{\rm d}N}{{\rm d}y} \;
\frac{p_{\rm T}(n-1)(n-2)}{nC[nC+m_{0}(n-2)]} \;
\left ( 1 + \frac{m_{\rm T}-m_{0}}{nC} \right )^{-n} \, ,
\end{equation}

where $m_{\rm T}=\sqrt{p_{\rm T}^2+m_{0}^2}$ is the transverse mass, $m_{0}$ is the rest mass of the particle (deuteron for the present analysis) and $n$, $C$ and \dndy~ are the free fit parameters.
As observed already in~\cite{ALICE:deuteronppPbPb2015} for inelastic collisions and in \cite{ALICE:longpaperpp2018} for light hadrons, the L\'evy-Tsallis function describes the spectra in all multiplicity classes rather well. 
The \pt-integrated yield per unit of rapidity (\dndy) at mid-rapidity and the mean transverse momentum \meanpt~are reported in Tab.~\ref{tab:fitSpectraValues}.
These are obtained by integrating the \pt-differential yields in the measured \pt~region and the fitted L\'evy-Tsallis function in the extrapolated regions at low and high \pt.
The fraction of yield contained in these two regions is also reported in the table.
The first uncertainty of \dndy~and \meanpt~reported in Tab.~\ref{tab:fitSpectraValues} represents the statistical uncertainty, whereas the second is the systematic uncertainty. 
The latter includes the uncertainty due to the extrapolation of the spectra, which amounts to about 4 to $9\%$ (from high to low multiplicity) of the integrated yield and to about 1 to $5\%$ of the mean~\pt. 
Both these estimates are derived by fitting the spectra with other functional forms, which describe the low and the high \pt~regions of the spectra in a different way. These include Boltzmann, Fermi-Dirac, Bose-Einstein, $m_{\rm T}$-exponential and \pt-exponential distributions \cite{PHENIX:2004}.

Table~\ref{tab:fitSpectraValues} shows that the yield of deuterons and anti-deuterons
increases with multiplicity, mirroring the fact that the number of constituent nucleons per event is also rising~\cite{ALICE:longpaperpp2018}.
The multiplicity dependence of the \meanpt~reflects the observed hardening of the deuteron and anti-deuteron spectra
from low to high multiplicity.

\begin{figure}[!h]
\begin{center}
\includegraphics[scale=0.75]{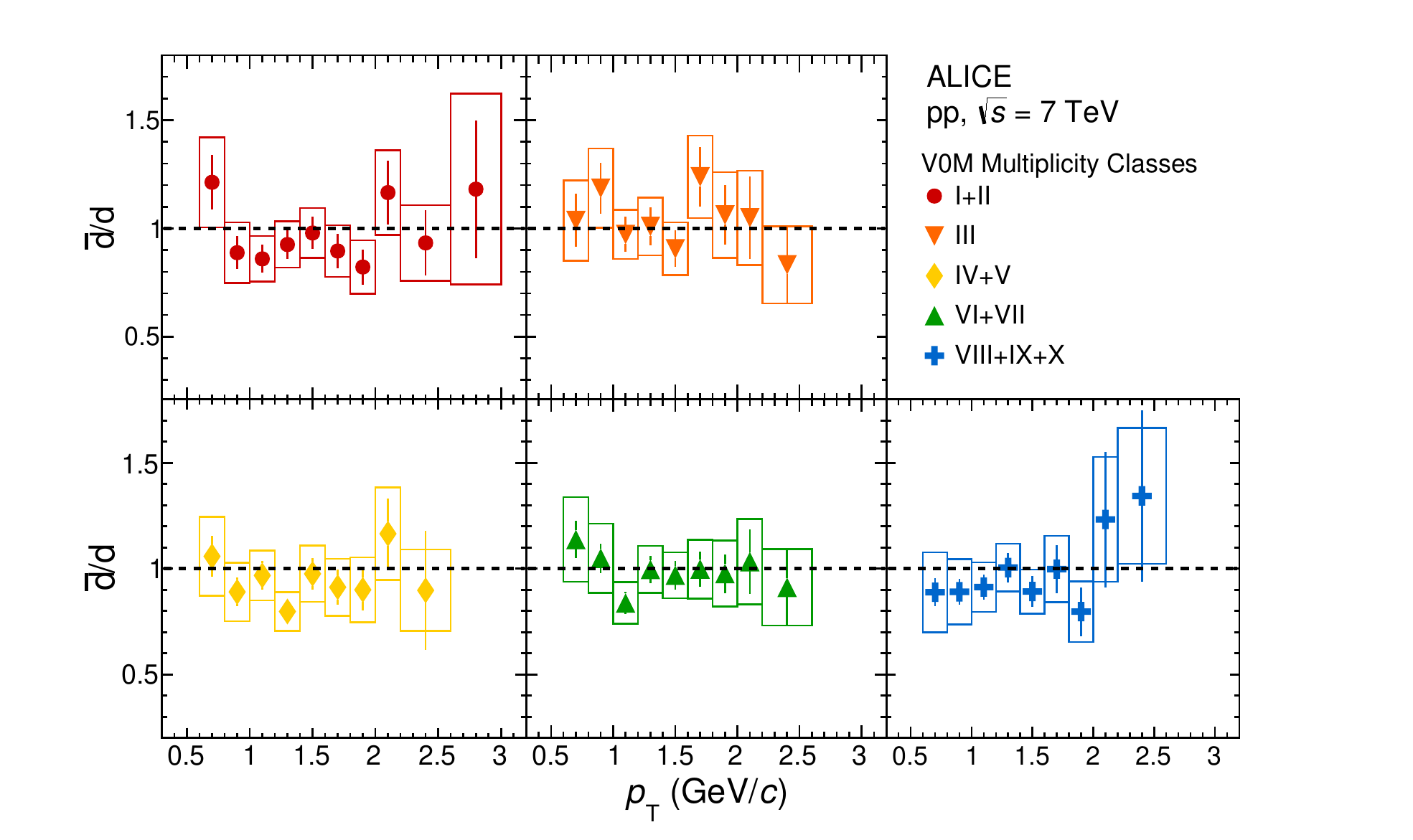}
\caption{Anti-deuteron to deuteron ratio as a function of \pt~in the considered multiplicity classes in pp collisions at \sqrtSE{7}.
The vertical bars represent the statistical uncertainty and the open boxes the systematic ones.} 
\label{fig:particlesRatios}
\end{center}
\end{figure}

The anti-deuteron to deuteron ratio is shown in Fig.~\ref{fig:particlesRatios} for the considered multiplicity classes.
These ratios are compatible with unity within $2\sigma$ (where $\sigma$ is the uncertainty in each \pt~bin) 
in the measured \pt~range and for all multiplicity classes, and are in agreement with results for protons~\cite{ALICE:longpaperpp2018}.
According to coalescence models, ${\rm \overline{d}}/{\rm d}$ is equal to $({\rm \overline{p}}/{\rm p})^{2}$ and 
the anti-proton to proton ratio is indeed compatible with unity~\cite{ALICE:longpaperpp2018}, 
independent of \pt~and of charged-particle multiplicity.
For each multiplicity class, the average of the anti-deuteron to deuteron ratio over all \pt~bins in Fig.~\ref{fig:particlesRatios} is reported in Tab.~\ref{tab:particlesRatios}.

\begin{table}[!t]
\begin{center}
\begin{tabular}{cc}
\hline
\hline
\rule{0pt}{2.5ex}
Multiplicity Class & ${\rm \overline{d}}/{\rm d}$\\
\hline
\rule{0pt}{2.5ex}
I+II & $ 0.93 \pm 0.03  \pm 0.13 $\\
III  & $ 1.01 \pm 0.04 \pm 0.15 $\\
IV+V & $ 0.92 \pm 0.03 \pm 0.13 $\\
VI+VII & $ 0.96 \pm 0.03 \pm 0.13 $\\ 
VIII+IX+X & $ 0.93 \pm 0.03 \pm 0.14 $\\
\hline
\hline
\end{tabular}
\medskip 
\caption{Anti-deuteron to deuteron ratio averaged over all measured \pt~bins in each multiplicity class in pp collisions at \sqrtSE{7}. The first uncertainty is statistical and the second is the systematic contribution.}
\label{tab:particlesRatios}
\end{center}
\end{table}

\subsection{Coalescence parameter \btwo} \label{sec:b2}
The production of light nuclei and anti-nuclei in pp collisions is expected to be the result of the coalescence of protons and neutrons
that are nearby in space and have similar velocities at the last stage of the collision. 
This process is described by models with the parameter $B_A$, where $A$ is the mass number of the nucleus under study.
Here, it corresponds to \btwo, which relates the invariant differential yield
of deuterons to the one of protons via the following equation \cite{Butler:1963,Scheibl:1998}

\begin{equation}
  \label{eq:B2}
  \frac{1}{2 \pi p_{\rm T}^{\rm d}} \frac{{\rm d}^{2}N^{\rm d}}{{\rm d}p_{\rm T}^{\rm d}{\rm d}y} = B_2 \;
  \left ( \frac{1}{2 \pi p_{\rm T}^{\rm p}} \frac{{\rm d}^{2}N^{\rm p}}{{\rm d}p_{\rm T}^{\rm p}{\rm d}y} \right )^2 \,.
\end{equation}

\begin{figure}[!h]
\begin{center}
\includegraphics[scale=0.55]{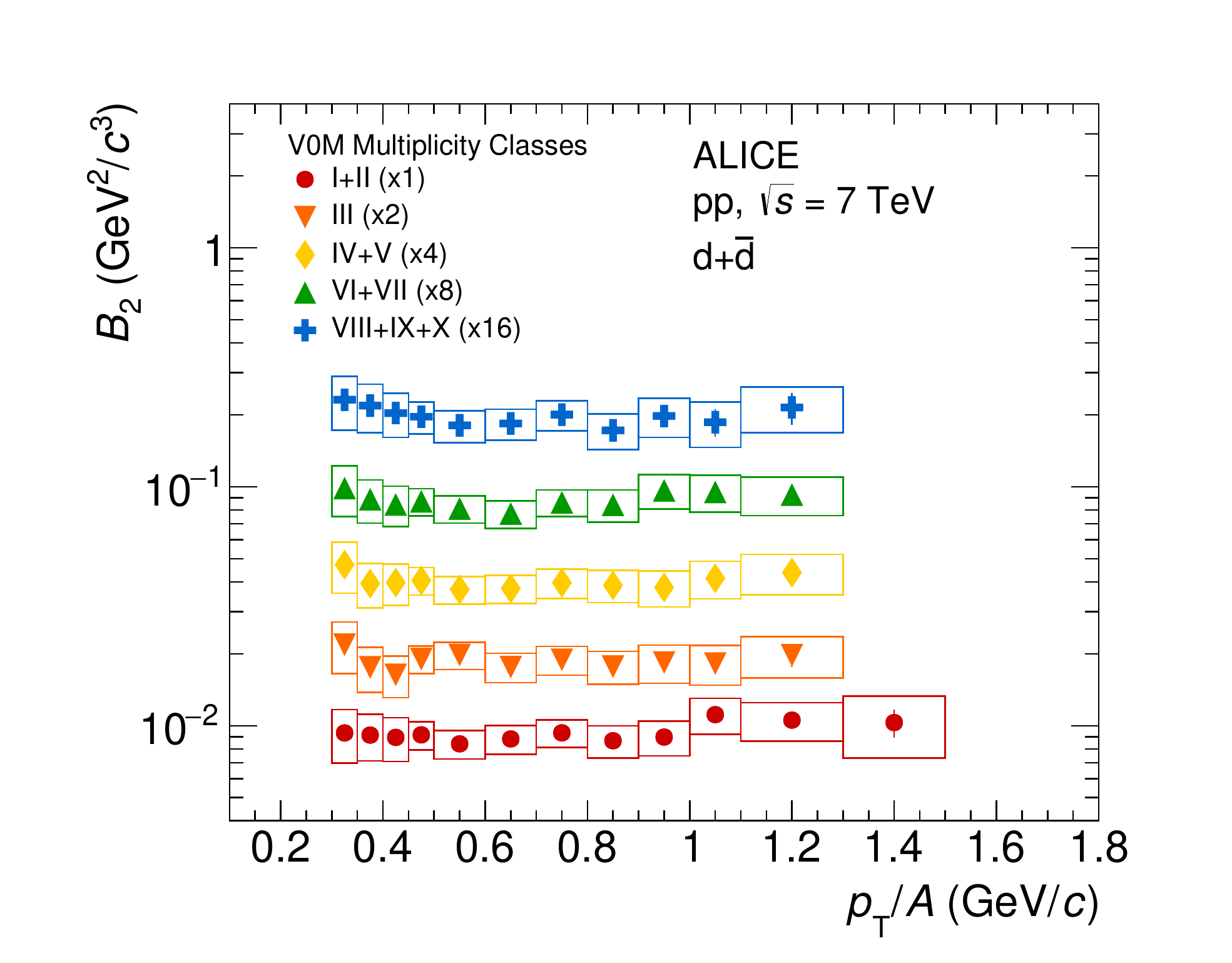}
\caption{Coalescence parameter \btwo~of (anti-)deuterons as a function of the transverse momentum per nucleon, \ptOa,
in the considered multiplicity classes in pp collisions at \sqrtSE{7}. The vertical bars represent the statistical uncertainties, the open boxes the systematic ones. The distributions in each class are scaled by constant factors to improve visibility.}
\label{fig:B2ddbar}
\end{center}
\end{figure}

In Eq.~\ref{eq:B2} the proton yield is measured at a value of half of the deuteron transverse momentum i.e. $p_{\rm T}^{\rm p}=p_{\rm T}^{\rm d}/2$
and neutrons are assumed to have the same invariant differential yield as protons.
Figure~\ref{fig:B2ddbar} shows the \btwo~parameter computed according to Eq.~\ref{eq:B2}
as a function of the transverse momentum per nucleon (\ptOa) for the different multiplicity classes, scaled by constant factors.
The differential yields for deuterons and anti-deuterons shown in Fig.~\ref{fig:spectra_ddbar} are used. The \pt~spectra of (anti-)protons are those published in~\cite{ALICE:longpaperpp2018}.
The statistical uncertainties in Fig.~\ref{fig:B2ddbar} are dominated by those of (anti-)deuterons,
while the systematic uncertainties by those of (anti-)protons, because the proton term enters to the square power in Eq.~\ref{eq:B2}.
In any of the considered multiplicity classes, within the experimental precision \btwo~does not show a significant \pt~dependence
as expected in a simple coalescence model~\cite{Butler:1963}, where a point-like source is assumed that emits nucleons without any correlation between proton and neutron momenta.

\begin{figure}[!h]
\begin{center}
\includegraphics[scale=0.6]{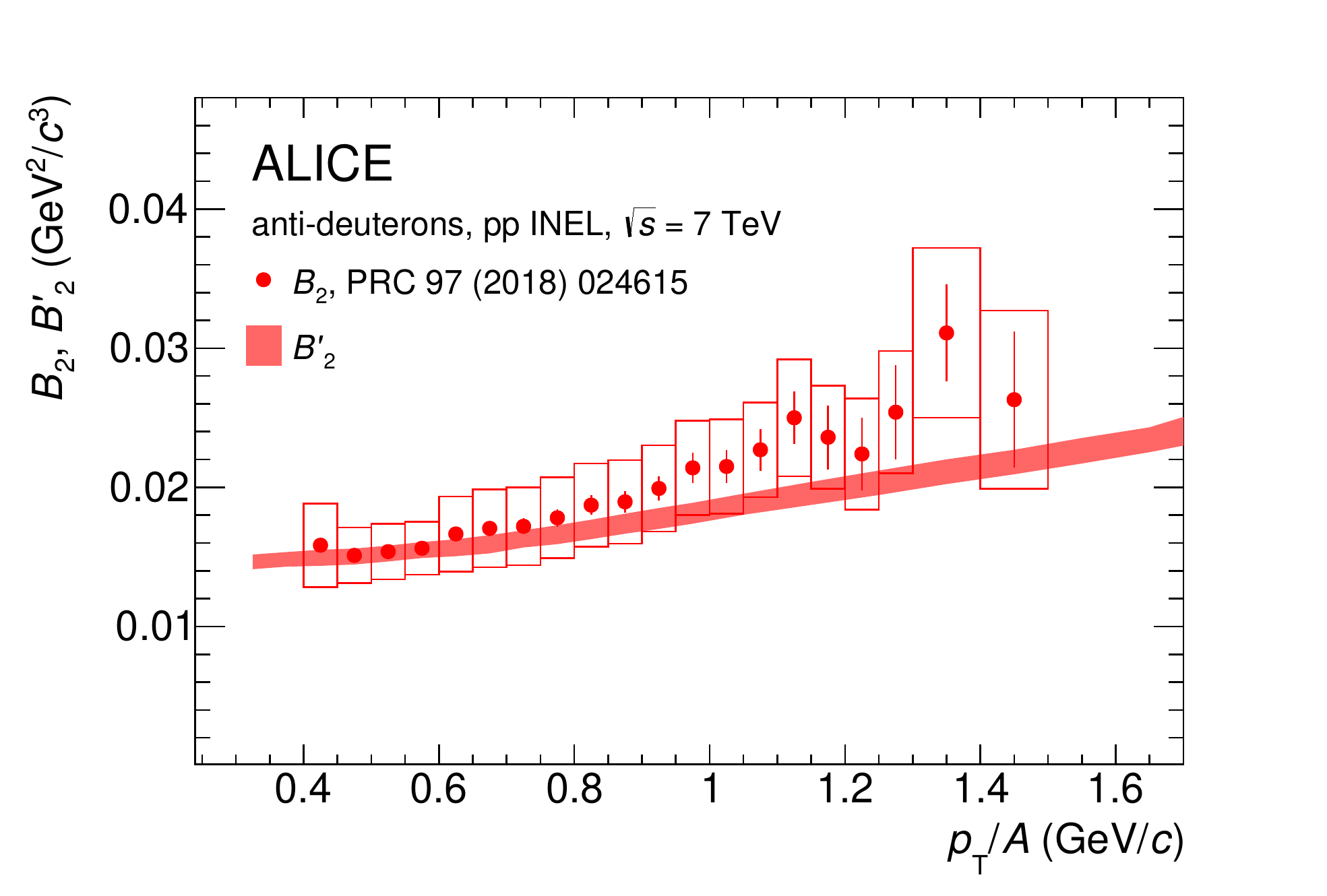}
\caption{Coalescence parameter $B_2^{\prime}$ of anti-deuterons as a function of the transverse momentum per nucleon \ptOa~(red shaded band, see text for details).
The result is compared with the experimental data for \btwo~measured in inelastic pp collisions at \sqrtSE{7}~\cite{ALICE:nucleipp2017}.}
\label{fig:B2coal}
\end{center}
\end{figure}

In \cite{ALICE:nucleipp2017}, where the results have been reported for inelastic pp collisions without any selection on the event multiplicity, the \btwo~parameter (red circles in Fig.~\ref{fig:B2coal}) was found to increase with the transverse momentum.
This trend was reproduced by an afterburner model~\cite{ALICE:afterburner2017}, which looks for correlations between nucleons produced by QCD-inspired event generators, and explained as a hard scattering effect~\cite{ALICE:nucleipp2017}.
In this work the coalescence parameter is re-evaluated for the multiplicity-integrated sample, indicated hereafter as $B_2^{\prime}$, by means of the following equation
\begin{equation}
  \label{eq:B2prime}‎‎
  B_2^{\prime}  =
  \frac { \sum\limits_{i={\rm I+II}}^{\rm VIII \; to \; X} \;\; (N_i / N) \; B_2^i \; (S_{{\rm p}}^i)^2 }
        { \left (\; \sum\limits_{i={\rm I+II}}^{\rm VIII \; to \; X} \;\; (N_i / N) \; S_{{\rm p}}^i \right )^2 } \,, 
\end{equation}
where $S_{\rm p}^i = 1/(2 \pi p_{\rm T}) {\rm d}^{2}N_{\rm p}^i/({\rm d}p_{\rm T}{\rm d}y)$ is the invariant differential yield of protons or anti-protons~\cite{ALICE:longpaperpp2018},
and $N_i / N$ the fraction of events in the $i$-th multiplicity class.
The set of the \pt-independent $B_2^i$ measured in this work are also used as inputs of Eq.~\ref{eq:B2prime}.
The result for \dbar~is shown in Fig.~\ref{fig:B2coal} as a red shaded band,
after being normalised to inelastic collisions via the scaling factor 0.852~\cite{ALICE:inelasticCrossSec2013}.
The width of the band represents an uncertainty of about 4$\%$.
This uncertainty includes a 2-3$\%$ contribution obtained by considering finer multiplicity classes than those used in the anti-deuteron analysis (anti-proton spectra are measured in \cite{ALICE:longpaperpp2018}, \btwo~has been interpolated), summed in quadrature to a 3$\%$ difference between deuteron and anti-deuteron results.
The level of agreement with the experimental points~from \cite{ALICE:nucleipp2017} indicates that part of the rise of \btwo, in the measured \ptOa~range, can be explained within a simple coalescence picture as a consequence of the hardening of the proton spectra with increasing multiplicity.
The hint for deviation at high \pt~leaves room for additional hard scattering effects, as the one invoked in~\cite{ALICE:nucleipp2017, ALICE:afterburner2017}.

It is worth noting that once the \btwo~parameter is measured directly from the multiplicity-integrated sample and normalised to inelastic collisions, the result obtained here is in agreement with the one published in \cite{ALICE:nucleipp2017}.
In central \PbPb~collisions the coalescence parameter exhibits an increasing trend with the transverse momentum~\cite{ALICE:deuteronppPbPb2015} that might be attributed to the presence of collective flow~\cite{Polleri:1997}. 

\begin{figure}[!h]
\begin{center}
\includegraphics[scale=0.7]{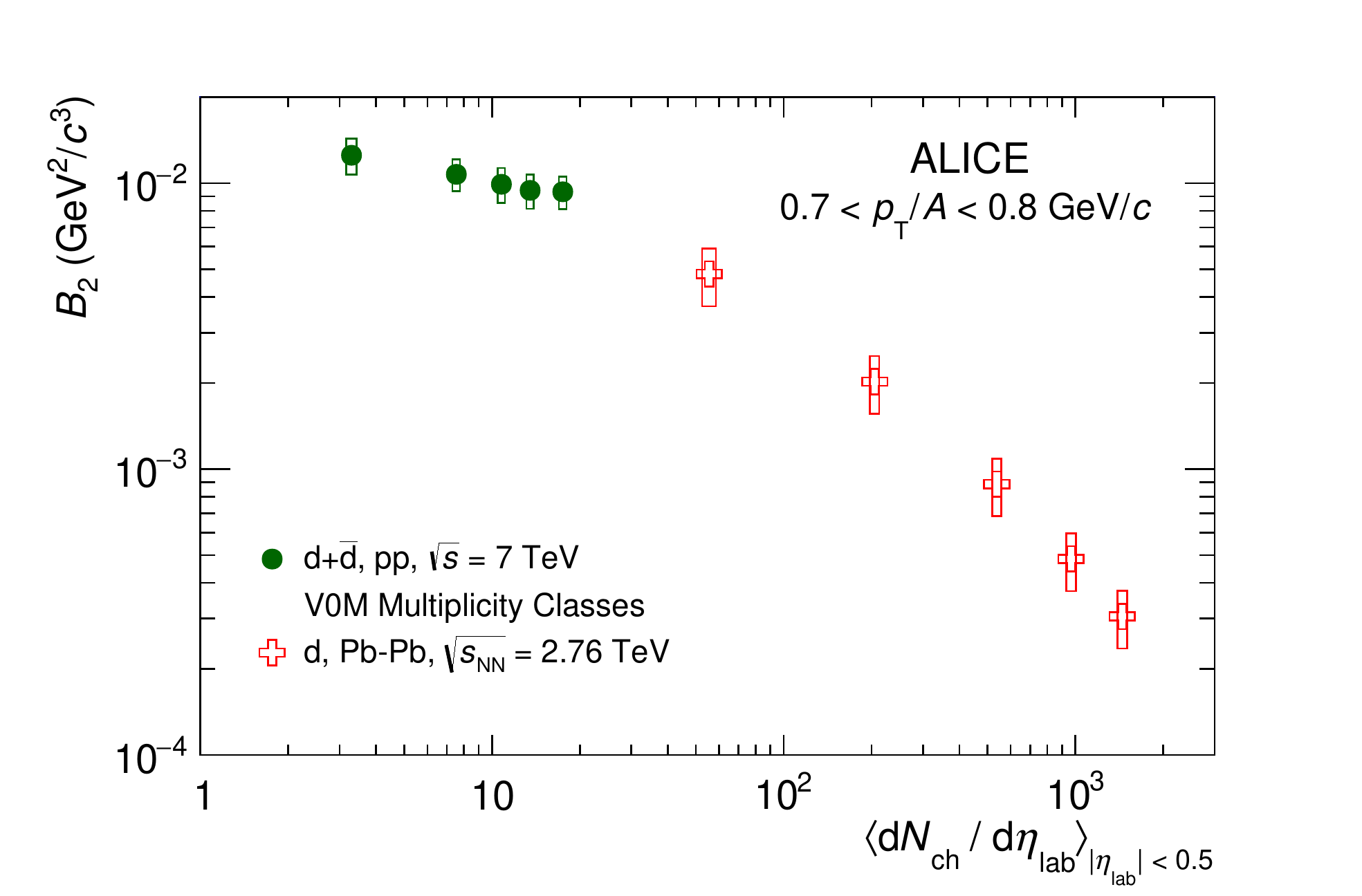}
\caption{Coalescence parameter \btwo~of (anti-)deuterons as a function of charged-particle multiplicity at mid-rapidity
in pp and \PbPb~collisions~\cite{ALICE:deuteronppPbPb2015} at the LHC at the transverse momentum per nucleon of 0.7~\textless~\ptOa~\textless~0.8~\GeVc.
The open boxes represent the systematic uncertainties.}
\label{fig:B2vsColls}
\end{center}
\end{figure}

The \btwo~parameter for one selected interval of transverse momentum per nucleon (0.7~\textless~\ptOa~\textless~0.8~\GeVc) is shown in Fig.~\ref{fig:B2vsColls} as a function of charged-particle multiplicity density at mid-rapidity and compared to the measurements in \PbPb~collisions at \sqrtSnnE{2.76}~\cite{ALICE:deuteronppPbPb2015}.
In a simple coalescence model~\cite{Butler:1963,Csernai:1986}, the \btwo~parameter is expected to be dependent only on the maximum relative momentum of the constituent nucleons coalescing in the bound state and therefore no multiplicity dependence is predicted.
In pp collisions (dark green circles in Fig.~\ref{fig:B2vsColls}), the extracted \btwo~is observed to vary by about 25$\%$ from the lowest to the highest multiplicity reached in the present analysis.
This effect is more pronounced in \PbPb~collisions and suggests that the increasing volume of the particle-emitting source -- which reduces the coalescence probability -- has to be taken into account, as done in more elaborate coalescence models~\cite{Scheibl:1998}. 


\begin{figure}[!h]
\begin{center}
\includegraphics[scale=0.65]{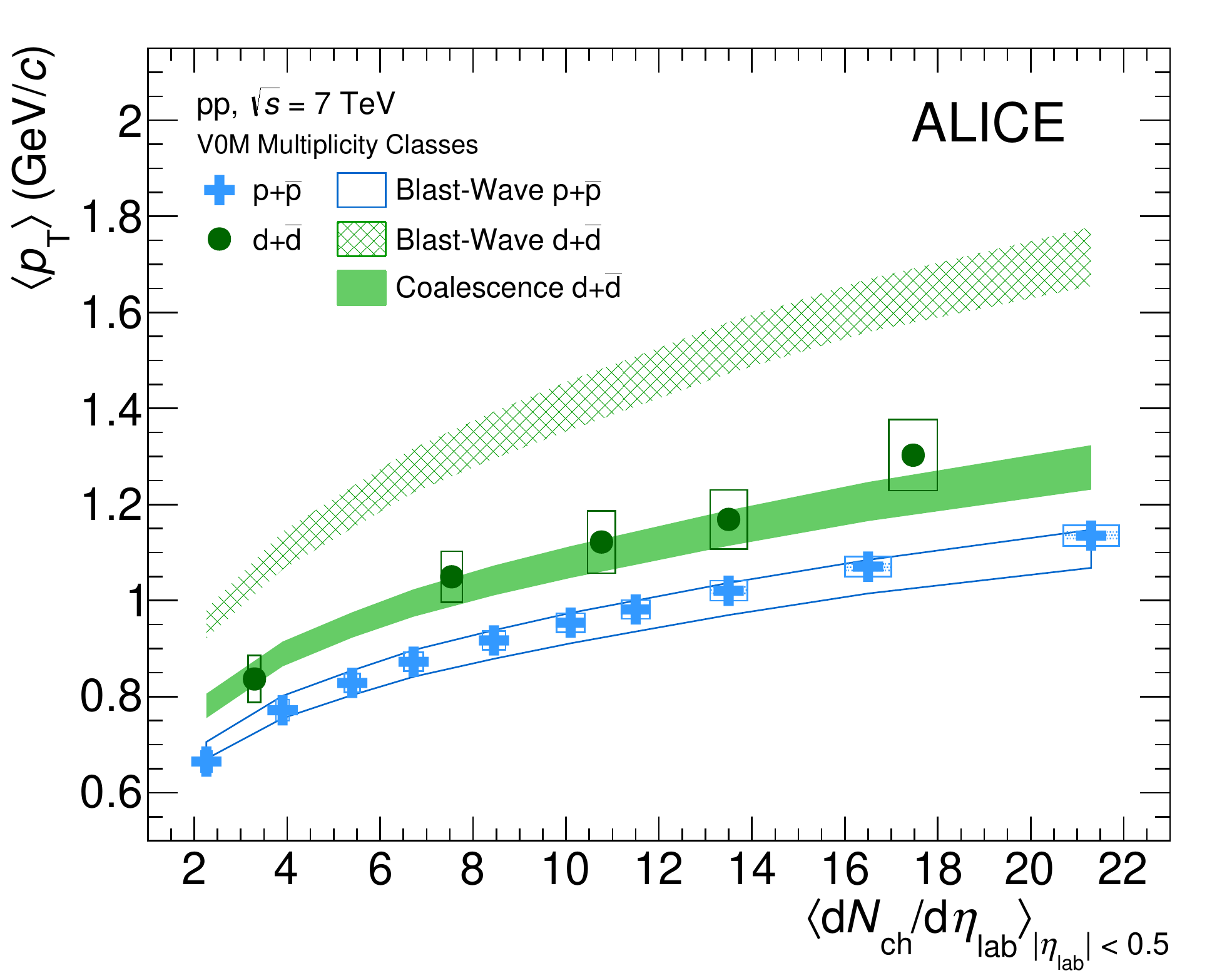}
\caption{Mean transverse momentum \meanpt~of deuterons and protons as a function of charged-particle multiplicity at mid-rapidity
in pp collisions at the LHC.
The open boxes represent the total systematic uncertainty while the contribution that is uncorrelated across multiplicity (where estimated) is shown with the shaded boxes. 
The full shaded area corresponds to the expected mean \pt~of deuterons from a simple coalescence model assuming a \pt-independent \btwo~value.
The hollow and dashed areas correspond to the mean \pt~of protons and deuterons calculated by using the Blast-Wave parameters
that simultaneously fit to the pion, kaon and proton spectra.}
\label{fig:meanpT}
\end{center}
\end{figure}

\subsection{Mean transverse momentum}
The mean transverse momenta of deuterons and protons are shown as a function of the charged-particle multiplicity in pp collisions in Fig.~\ref{fig:meanpT}.
The difference between deuteron and proton mean momenta is significant, except at extremely low charged-particle multiplicity.
In high-multiplicity pp collisions, the ratio between the \meanpt~of deuterons and protons is about 1.2 and is smaller than the value (about 1.6) measured in central \PbPb~collisions~\cite{ALICE:deuteronppPbPb2015},
where the established mass ordering is in general attributed
to the emission of particles from a radially expanding source.\\
In pp collisions the multiplicity dependence of the deuteron mean transverse momentum is well reproduced by computing the deuteron spectra using Eq.~\ref{eq:B2}
with the proton spectra as input and assuming, as in a simple coalescence model, a \pt-independent \btwo~value.
Note that in central \PbPb~collisions the Blast-Wave model~\cite{BW:1993}
-- a hydrodynamic-inspired model which describes particle production assuming that these are emitted from an expanding thermalised source -- simultaneously fits light nuclei (deuterons and $^3{\rm He}$) together with light hadrons~\cite{ALICE:deuteronppPbPb2015}. 
On the contrary, in pp collisions, the \meanpt~of deuterons is not correctly reproduced
by using the Blast-Wave parameters that simultaneously describe pion, kaon and proton spectra
from \cite{ALICE:longpaperpp2018}, as clearly shown in Fig.~\ref{fig:meanpT}.
Since the Blast-Wave model is able to reproduce experimental data solely in \PbPb~collisions,
we have evidence that a full hydrodynamic approach does not concurrently describe the production of light hadrons and nuclei in pp collisions.
The latter is consistent with a coalescence picture where the formation of weakly bound composite particles is expected to 
occur only at the last stage of the system evolution after the collision, namely after the kinetic freeze-out.

\subsection{Deuteron-to-proton ratio}
\begin{figure}[!h]
\begin{center}
\includegraphics[scale=0.7]{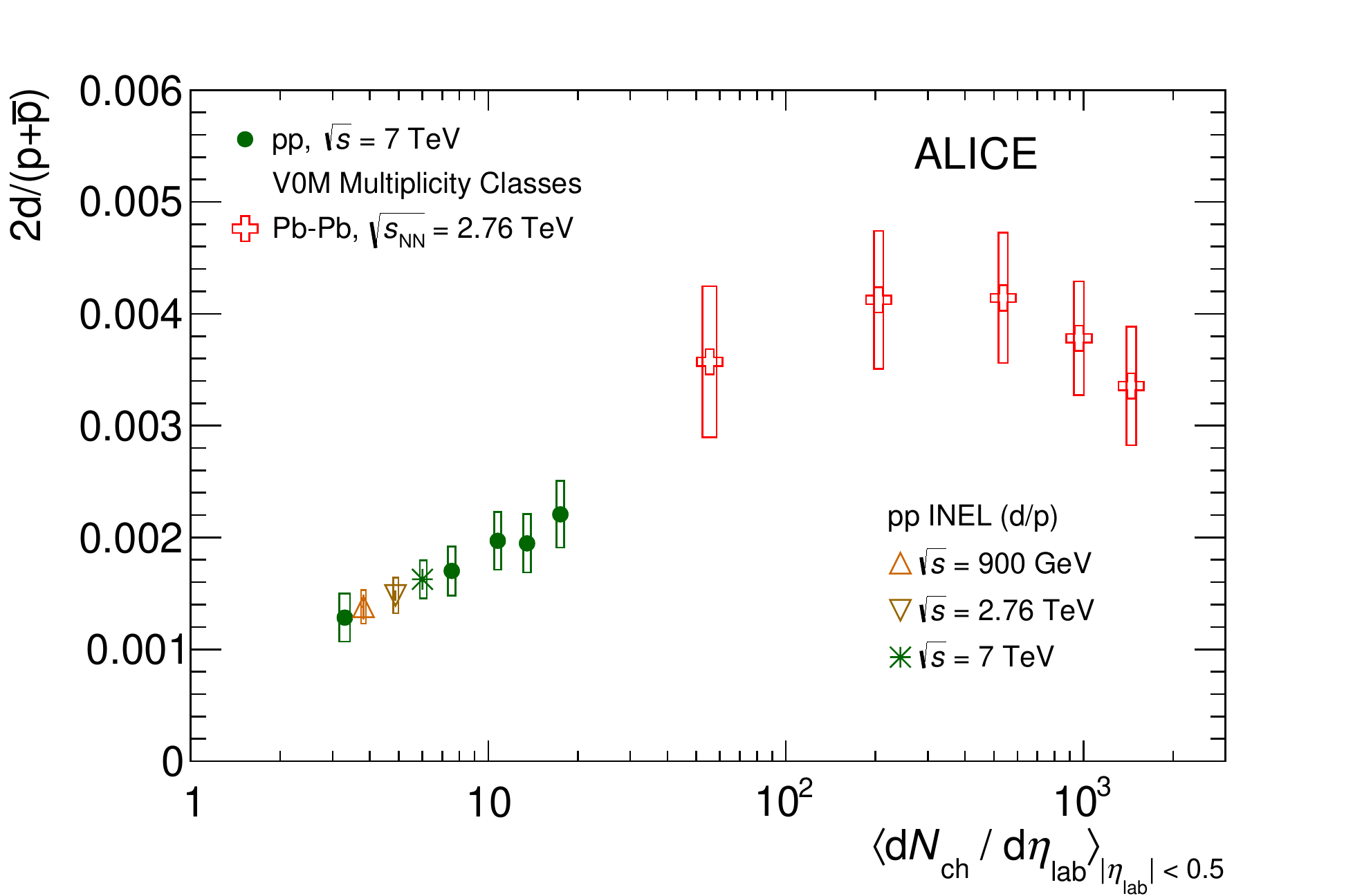}
\caption{Ratio between the \pt-integrated yield of deuterons and protons as a function 
of charged-particle multiplicity at mid-rapidity in pp (this work) and \PbPb~collisions~\cite{ALICE:deuteronppPbPb2015} at the LHC.
The deuteron-to-proton ratio measured in inelastic pp collisions at \sqrtSE{0.9, 2.76~${\rm and}$~7}~\cite{ALICE:nucleipp2017} has also been reported.
}
\label{fig:dp}
\end{center}
\end{figure}

Figure~\ref{fig:dp} shows the ratio between the \pt-integrated yield of deuterons and protons
as a function of multiplicity, including all the presently available measurements performed at the LHC.
For computing the multiplicity-dependent ratio in pp collisions at \sqrtSE{7}, the deuteron yields reported in Tab.~\ref{tab:fitSpectraValues} are used.
The \dndy~of protons are those reported in \cite{ALICE:npstrange2017}.
In a naive approach, one would predict an increase of the deuteron-to-proton ratio since the number of nucleons increases with the charged-particle multiplicity. In pp collisions, the observed trend of the d/p ratio is in qualitative agreement with this expectation, further supported by the fact that the systematic uncertainties are expected to be largely correlated across multiplicity.
In more sophisticated coalescence models~\cite{Scheibl:1998}, the source volume is also taken into account and
the rise of the d/p ratio is expected to be the result of an enhanced nucleon density, and not simply related to the nucleon abundances.
The prediction of~\cite{Scheibl:1998} qualitatively describes the data if the rise in the nucleon abundance dominates over the increase in the volume size in pp collisions. 
No significant multiplicity dependence of the d/p ratio is observed in \PbPb~collisions within the achieved experimental precision~\cite{ALICE:deuteronppPbPb2015},
in agreement with expectations from thermal-statistical models~\cite{Andronic:nuclei2011, Cleymans:2011}.

\section{Conclusions}
The transverse-momentum spectra of deuterons and anti-deuterons in pp collisions at $\sqrt{s}$~=~7~TeV have been presented in five multiplicity classes.
They are combined with the primary proton spectra to extract the coalescence parameter \btwo.
The latter~exhibits an approximately constant behaviour with the transverse momentum per nucleon in multiplicity classes in the measured \ptOa~range, in agreement with a simple coalescence model, where uncorrelated particle emission from a point-like source is assumed.
A simple coalescence picture cannot, however, explain the multiplicity dependence of the \btwo~parameter at fixed transverse momentum (\ptOa~= 0.75 \GeVc), observed also in \PbPb~collisions. 
Instead, these observations point toward a dependence of the coalescence process on the volume of the particle-emitting source.
In fact, the increasing volume of the particle-emitting source with multiplicity plays an effective role in reducing the coalescence probability as predicted by more elaborate models. 
These models are able to describe data even in the smallest colliding system at the LHC, as reported in this letter, where the spatial extension of the source is comparable to the deuteron size.
Coalescence model calculations, precisely correlating the size of the hadronic emission region with the multiplicity, need to be performed to quantitatively support the current interpretation of the results.

The mean transverse momentum of deuterons has been measured as a function of the charged-particle multiplicity.
In pp collisions, the hydrodynamic-inspired Blast-Wave model, which assumes that the particles are emitted thermally from an expanding source, does not describe the production of nuclei with identical freeze-out conditions as lighter hadrons.
While in central \PbPb~collisions there is evidence that nuclei and anti-nuclei participate in the expansion of the fireball together with non-composite light hadrons, in \pp~collisions such evidence is missing.

All presently available measurements of the \pt-integrated d/p ratio at the LHC have been discussed as a function of the charged-particle multiplicity.
The observed multiplicity dependence of the d$/$p ratio 
suggests that the rise with multiplicity of the number of nucleons available for coalescence is faster than the increase of the source volume in small colliding systems at the LHC. 
The multiplicity dependence of d$/$p, as well as that of \btwo, hints at a continuous evolution of deuteron production from low-multiplicity pp to \PbPb~collisions.
Measurements at intermediate multiplicities, such as those reached in \pPb~collisions, are being performed to confirm this picture.

The observed similarities between \pp~and heavy-ion collisions
can be traced back to common underlying production mechanisms of light (anti-)nuclei. 
The differences, such as the one appearing in the mean transverse momentum of deuterons, are extremely interesting because they can shed light on the possibility that nuclei may emerge at different stages of the collision depending on the initial conditions.

%
%
\newenvironment{acknowledgement}{\relax}{\relax}
\begin{acknowledgement}
\section*{Acknowledgements}

The ALICE Collaboration would like to thank all its engineers and technicians for their invaluable contributions to the construction of the experiment and the CERN accelerator teams for the outstanding performance of the LHC complex.
The ALICE Collaboration gratefully acknowledges the resources and support provided by all Grid centres and the Worldwide LHC Computing Grid (WLCG) collaboration.
The ALICE Collaboration acknowledges the following funding agencies for their support in building and running the ALICE detector:
A. I. Alikhanyan National Science Laboratory (Yerevan Physics Institute) Foundation (ANSL), State Committee of Science and World Federation of Scientists (WFS), Armenia;
Austrian Academy of Sciences, Austrian Science Fund (FWF): [M 2467-N36] and Nationalstiftung f\"{u}r Forschung, Technologie und Entwicklung, Austria;
Ministry of Communications and High Technologies, National Nuclear Research Center, Azerbaijan;
Conselho Nacional de Desenvolvimento Cient\'{\i}fico e Tecnol\'{o}gico (CNPq), Universidade Federal do Rio Grande do Sul (UFRGS), Financiadora de Estudos e Projetos (Finep) and Funda\c{c}\~{a}o de Amparo \`{a} Pesquisa do Estado de S\~{a}o Paulo (FAPESP), Brazil;
Ministry of Science \& Technology of China (MSTC), National Natural Science Foundation of China (NSFC) and Ministry of Education of China (MOEC) , China;
Croatian Science Foundation and Ministry of Science and Education, Croatia;
Centro de Aplicaciones Tecnol\'{o}gicas y Desarrollo Nuclear (CEADEN), Cubaenerg\'{\i}a, Cuba;
Ministry of Education, Youth and Sports of the Czech Republic, Czech Republic;
The Danish Council for Independent Research | Natural Sciences, the Carlsberg Foundation and Danish National Research Foundation (DNRF), Denmark;
Helsinki Institute of Physics (HIP), Finland;
Commissariat \`{a} l'Energie Atomique (CEA), Institut National de Physique Nucl\'{e}aire et de Physique des Particules (IN2P3) and Centre National de la Recherche Scientifique (CNRS) and Rl\'{e}gion des  Pays de la Loire, France;
Bundesministerium f\"{u}r Bildung, Wissenschaft, Forschung und Technologie (BMBF) and GSI Helmholtzzentrum f\"{u}r Schwerionenforschung GmbH, Germany;
General Secretariat for Research and Technology, Ministry of Education, Research and Religions, Greece;
National Research, Development and Innovation Office, Hungary;
Department of Atomic Energy Government of India (DAE), Department of Science and Technology, Government of India (DST), University Grants Commission, Government of India (UGC) and Council of Scientific and Industrial Research (CSIR), India;
Indonesian Institute of Science, Indonesia;
Centro Fermi - Museo Storico della Fisica e Centro Studi e Ricerche Enrico Fermi and Istituto Nazionale di Fisica Nucleare (INFN), Italy;
Institute for Innovative Science and Technology , Nagasaki Institute of Applied Science (IIST), Japan Society for the Promotion of Science (JSPS) KAKENHI and Japanese Ministry of Education, Culture, Sports, Science and Technology (MEXT), Japan;
Consejo Nacional de Ciencia (CONACYT) y Tecnolog\'{i}a, through Fondo de Cooperaci\'{o}n Internacional en Ciencia y Tecnolog\'{i}a (FONCICYT) and Direcci\'{o}n General de Asuntos del Personal Academico (DGAPA), Mexico;
Nederlandse Organisatie voor Wetenschappelijk Onderzoek (NWO), Netherlands;
The Research Council of Norway, Norway;
Commission on Science and Technology for Sustainable Development in the South (COMSATS), Pakistan;
Pontificia Universidad Cat\'{o}lica del Per\'{u}, Peru;
Ministry of Science and Higher Education and National Science Centre, Poland;
Korea Institute of Science and Technology Information and National Research Foundation of Korea (NRF), Republic of Korea;
Ministry of Education and Scientific Research, Institute of Atomic Physics and Ministry of Research and Innovation and Institute of Atomic Physics, Romania;
Joint Institute for Nuclear Research (JINR), Ministry of Education and Science of the Russian Federation, National Research Centre Kurchatov Institute, Russian Science Foundation and Russian Foundation for Basic Research, Russia;
Ministry of Education, Science, Research and Sport of the Slovak Republic, Slovakia;
National Research Foundation of South Africa, South Africa;
Swedish Research Council (VR) and Knut \& Alice Wallenberg Foundation (KAW), Sweden;
European Organization for Nuclear Research, Switzerland;
National Science and Technology Development Agency (NSDTA), Suranaree University of Technology (SUT) and Office of the Higher Education Commission under NRU project of Thailand, Thailand;
Turkish Atomic Energy Agency (TAEK), Turkey;
National Academy of  Sciences of Ukraine, Ukraine;
Science and Technology Facilities Council (STFC), United Kingdom;
National Science Foundation of the United States of America (NSF) and United States Department of Energy, Office of Nuclear Physics (DOE NP), United States of America.    
\end{acknowledgement}

\bibliographystyle{utphys}   
\bibliography{biblio}

\newpage
\appendix
\section{The ALICE Collaboration}
\label{app:collab}

\begingroup
\small
\begin{flushleft}
S.~Acharya\Irefn{org140}\And 
F.T.-.~Acosta\Irefn{org20}\And 
D.~Adamov\'{a}\Irefn{org93}\And 
S.P.~Adhya\Irefn{org140}\And 
A.~Adler\Irefn{org74}\And 
J.~Adolfsson\Irefn{org80}\And 
M.M.~Aggarwal\Irefn{org98}\And 
G.~Aglieri Rinella\Irefn{org34}\And 
M.~Agnello\Irefn{org31}\And 
Z.~Ahammed\Irefn{org140}\And 
S.~Ahmad\Irefn{org17}
% ...
\And
\renewcommand\labelenumi{\textsuperscript{\theenumi}~}

\section*{Affiliation notes}
\renewcommand\theenumi{\roman{enumi}}
\begin{Authlist}
\item \Adef{org*}Deceased
\item \Adef{orgI}Dipartimento DET del Politecnico di Torino, Turin, Italy
\item \Adef{orgII}M.V. Lomonosov Moscow State University, D.V. Skobeltsyn Institute of Nuclear, Physics, Moscow, Russia
\item \Adef{orgIII}Department of Applied Physics, Aligarh Muslim University, Aligarh, India
\item \Adef{orgIV}Institute of Theoretical Physics, University of Wroclaw, Poland
\end{Authlist}

\section*{Collaboration Institutes}
\renewcommand\theenumi{\arabic{enumi}~}
\begin{Authlist}
\item \Idef{org1}A.I. Alikhanyan National Science Laboratory (Yerevan Physics Institute) Foundation, Yerevan, Armenia
\item \Idef{org2}Bogolyubov Institute for Theoretical Physics, National Academy of Sciences of Ukraine, Kiev, Ukraine
\item \Idef{org3}Bose Institute, Department of Physics  and Centre for Astroparticle Physics and Space Science (CAPSS), Kolkata, India
\item \Idef{org4}Budker Institute for Nuclear Physics, Novosibirsk, Russia
\item \Idef{org5}California Polytechnic State University, San Luis Obispo, California, United States
\item \Idef{org6}Central China Normal University, Wuhan, China
\item \Idef{org7}Centre de Calcul de l'IN2P3, Villeurbanne, Lyon, France
\item \Idef{org8}Centro de Aplicaciones Tecnol\'{o}gicas y Desarrollo Nuclear (CEADEN), Havana, Cuba
\item \Idef{org9}Centro de Investigaci\'{o}n y de Estudios Avanzados (CINVESTAV), Mexico City and M\'{e}rida, Mexico
\item \Idef{org10}Centro Fermi - Museo Storico della Fisica e Centro Studi e Ricerche ``Enrico Fermi', Rome, Italy
\item \Idef{org11}Chicago State University, Chicago, Illinois, United States
\item \Idef{org12}China Institute of Atomic Energy, Beijing, China
\item \Idef{org13}Chonbuk National University, Jeonju, Republic of Korea
\item \Idef{org14}Comenius University Bratislava, Faculty of Mathematics, Physics and Informatics, Bratislava, Slovakia
\item \Idef{org15}COMSATS Institute of Information Technology (CIIT), Islamabad, Pakistan
\item \Idef{org16}Creighton University, Omaha, Nebraska, United States
\item \Idef{org17}Department of Physics, Aligarh Muslim University, Aligarh, India
\item \Idef{org18}Department of Physics, Pusan National University, Pusan, Republic of Korea
\item \Idef{org19}Department of Physics, Sejong University, Seoul, Republic of Korea
\item \Idef{org20}Department of Physics, University of California, Berkeley, California, United States
\item \Idef{org21}Department of Physics, University of Oslo, Oslo, Norway
\item \Idef{org22}Department of Physics and Technology, University of Bergen, Bergen, Norway
\item \Idef{org23}Dipartimento di Fisica dell'Universit\`{a} 'La Sapienza' and Sezione INFN, Rome, Italy
\item \Idef{org24}Dipartimento di Fisica dell'Universit\`{a} and Sezione INFN, Cagliari, Italy
\item \Idef{org25}Dipartimento di Fisica dell'Universit\`{a} and Sezione INFN, Trieste, Italy
\item \Idef{org26}Dipartimento di Fisica dell'Universit\`{a} and Sezione INFN, Turin, Italy
\item \Idef{org27}Dipartimento di Fisica e Astronomia dell'Universit\`{a} and Sezione INFN, Bologna, Italy
\item \Idef{org28}Dipartimento di Fisica e Astronomia dell'Universit\`{a} and Sezione INFN, Catania, Italy
\item \Idef{org29}Dipartimento di Fisica e Astronomia dell'Universit\`{a} and Sezione INFN, Padova, Italy
\item \Idef{org30}Dipartimento di Fisica `E.R.~Caianiello' dell'Universit\`{a} and Gruppo Collegato INFN, Salerno, Italy
\item \Idef{org31}Dipartimento DISAT del Politecnico and Sezione INFN, Turin, Italy
\item \Idef{org32}Dipartimento di Scienze e Innovazione Tecnologica dell'Universit\`{a} del Piemonte Orientale and INFN Sezione di Torino, Alessandria, Italy
\item \Idef{org33}Dipartimento Interateneo di Fisica `M.~Merlin' and Sezione INFN, Bari, Italy
\item \Idef{org34}European Organization for Nuclear Research (CERN), Geneva, Switzerland
\item \Idef{org35}Faculty of Electrical Engineering, Mechanical Engineering and Naval Architecture, University of Split, Split, Croatia
\item \Idef{org36}Faculty of Engineering and Science, Western Norway University of Applied Sciences, Bergen, Norway
\item \Idef{org37}Faculty of Nuclear Sciences and Physical Engineering, Czech Technical University in Prague, Prague, Czech Republic
\item \Idef{org38}Faculty of Science, P.J.~\v{S}af\'{a}rik University, Ko\v{s}ice, Slovakia
\item \Idef{org39}Frankfurt Institute for Advanced Studies, Johann Wolfgang Goethe-Universit\"{a}t Frankfurt, Frankfurt, Germany
\item \Idef{org40}Gangneung-Wonju National University, Gangneung, Republic of Korea
\item \Idef{org41}Gauhati University, Department of Physics, Guwahati, India
\item \Idef{org42}Helmholtz-Institut f\"{u}r Strahlen- und Kernphysik, Rheinische Friedrich-Wilhelms-Universit\"{a}t Bonn, Bonn, Germany
\item \Idef{org43}Helsinki Institute of Physics (HIP), Helsinki, Finland
\item \Idef{org44}High Energy Physics Group,  Universidad Aut\'{o}noma de Puebla, Puebla, Mexico
\item \Idef{org45}Hiroshima University, Hiroshima, Japan
\item \Idef{org46}Hochschule Worms, Zentrum  f\"{u}r Technologietransfer und Telekommunikation (ZTT), Worms, Germany
\item \Idef{org47}Horia Hulubei National Institute of Physics and Nuclear Engineering, Bucharest, Romania
\item \Idef{org48}Indian Institute of Technology Bombay (IIT), Mumbai, India
\item \Idef{org49}Indian Institute of Technology Indore, Indore, India
\item \Idef{org50}Indonesian Institute of Sciences, Jakarta, Indonesia
\item \Idef{org51}INFN, Laboratori Nazionali di Frascati, Frascati, Italy
\item \Idef{org52}INFN, Sezione di Bari, Bari, Italy
\item \Idef{org53}INFN, Sezione di Bologna, Bologna, Italy
\item \Idef{org54}INFN, Sezione di Cagliari, Cagliari, Italy
\item \Idef{org55}INFN, Sezione di Catania, Catania, Italy
\item \Idef{org56}INFN, Sezione di Padova, Padova, Italy
\item \Idef{org57}INFN, Sezione di Roma, Rome, Italy
\item \Idef{org58}INFN, Sezione di Torino, Turin, Italy
\item \Idef{org59}INFN, Sezione di Trieste, Trieste, Italy
\item \Idef{org60}Inha University, Incheon, Republic of Korea
\item \Idef{org61}Institut de Physique Nucl\'{e}aire d'Orsay (IPNO), Institut National de Physique Nucl\'{e}aire et de Physique des Particules (IN2P3/CNRS), Universit\'{e} de Paris-Sud, Universit\'{e} Paris-Saclay, Orsay, France
\item \Idef{org62}Institute for Nuclear Research, Academy of Sciences, Moscow, Russia
\item \Idef{org63}Institute for Subatomic Physics, Utrecht University/Nikhef, Utrecht, Netherlands
\item \Idef{org64}Institute for Theoretical and Experimental Physics, Moscow, Russia
\item \Idef{org65}Institute of Experimental Physics, Slovak Academy of Sciences, Ko\v{s}ice, Slovakia
\item \Idef{org66}Institute of Physics, Homi Bhabha National Institute, Bhubaneswar, India
\item \Idef{org67}Institute of Physics of the Czech Academy of Sciences, Prague, Czech Republic
\item \Idef{org68}Institute of Space Science (ISS), Bucharest, Romania
\item \Idef{org69}Institut f\"{u}r Kernphysik, Johann Wolfgang Goethe-Universit\"{a}t Frankfurt, Frankfurt, Germany
\item \Idef{org70}Instituto de Ciencias Nucleares, Universidad Nacional Aut\'{o}noma de M\'{e}xico, Mexico City, Mexico
\item \Idef{org71}Instituto de F\'{i}sica, Universidade Federal do Rio Grande do Sul (UFRGS), Porto Alegre, Brazil
\item \Idef{org72}Instituto de F\'{\i}sica, Universidad Nacional Aut\'{o}noma de M\'{e}xico, Mexico City, Mexico
\item \Idef{org73}iThemba LABS, National Research Foundation, Somerset West, South Africa
\item \Idef{org74}Johann-Wolfgang-Goethe Universit\"{a}t Frankfurt Institut f\"{u}r Informatik, Fachbereich Informatik und Mathematik, Frankfurt, Germany
\item \Idef{org75}Joint Institute for Nuclear Research (JINR), Dubna, Russia
\item \Idef{org76}Korea Institute of Science and Technology Information, Daejeon, Republic of Korea
\item \Idef{org77}KTO Karatay University, Konya, Turkey
\item \Idef{org78}Laboratoire de Physique Subatomique et de Cosmologie, Universit\'{e} Grenoble-Alpes, CNRS-IN2P3, Grenoble, France
\item \Idef{org79}Lawrence Berkeley National Laboratory, Berkeley, California, United States
\item \Idef{org80}Lund University Department of Physics, Division of Particle Physics, Lund, Sweden
\item \Idef{org81}Nagasaki Institute of Applied Science, Nagasaki, Japan
\item \Idef{org82}Nara Women{'}s University (NWU), Nara, Japan
\item \Idef{org83}National and Kapodistrian University of Athens, School of Science, Department of Physics , Athens, Greece
\item \Idef{org84}National Centre for Nuclear Research, Warsaw, Poland
\item \Idef{org85}National Institute of Science Education and Research, Homi Bhabha National Institute, Jatni, India
\item \Idef{org86}National Nuclear Research Center, Baku, Azerbaijan
\item \Idef{org87}National Research Centre Kurchatov Institute, Moscow, Russia
\item \Idef{org88}Niels Bohr Institute, University of Copenhagen, Copenhagen, Denmark
\item \Idef{org89}Nikhef, National institute for subatomic physics, Amsterdam, Netherlands
\item \Idef{org90}NRC Kurchatov Institute IHEP, Protvino, Russia
\item \Idef{org91}NRNU Moscow Engineering Physics Institute, Moscow, Russia
\item \Idef{org92}Nuclear Physics Group, STFC Daresbury Laboratory, Daresbury, United Kingdom
\item \Idef{org93}Nuclear Physics Institute of the Czech Academy of Sciences, \v{R}e\v{z} u Prahy, Czech Republic
\item \Idef{org94}Oak Ridge National Laboratory, Oak Ridge, Tennessee, United States
\item \Idef{org95}Ohio State University, Columbus, Ohio, United States
\item \Idef{org96}Petersburg Nuclear Physics Institute, Gatchina, Russia
\item \Idef{org97}Physics department, Faculty of science, University of Zagreb, Zagreb, Croatia
\item \Idef{org98}Physics Department, Panjab University, Chandigarh, India
\item \Idef{org99}Physics Department, University of Jammu, Jammu, India
\item \Idef{org100}Physics Department, University of Rajasthan, Jaipur, India
\item \Idef{org101}Physikalisches Institut, Eberhard-Karls-Universit\"{a}t T\"{u}bingen, T\"{u}bingen, Germany
\item \Idef{org102}Physikalisches Institut, Ruprecht-Karls-Universit\"{a}t Heidelberg, Heidelberg, Germany
\item \Idef{org103}Physik Department, Technische Universit\"{a}t M\"{u}nchen, Munich, Germany
\item \Idef{org104}Research Division and ExtreMe Matter Institute EMMI, GSI Helmholtzzentrum f\"ur Schwerionenforschung GmbH, Darmstadt, Germany
\item \Idef{org105}Rudjer Bo\v{s}kovi\'{c} Institute, Zagreb, Croatia
\item \Idef{org106}Russian Federal Nuclear Center (VNIIEF), Sarov, Russia
\item \Idef{org107}Saha Institute of Nuclear Physics, Homi Bhabha National Institute, Kolkata, India
\item \Idef{org108}School of Physics and Astronomy, University of Birmingham, Birmingham, United Kingdom
\item \Idef{org109}Secci\'{o}n F\'{\i}sica, Departamento de Ciencias, Pontificia Universidad Cat\'{o}lica del Per\'{u}, Lima, Peru
\item \Idef{org110}Shanghai Institute of Applied Physics, Shanghai, China
\item \Idef{org111}St. Petersburg State University, St. Petersburg, Russia
\item \Idef{org112}Stefan Meyer Institut f\"{u}r Subatomare Physik (SMI), Vienna, Austria
\item \Idef{org113}SUBATECH, IMT Atlantique, Universit\'{e} de Nantes, CNRS-IN2P3, Nantes, France
\item \Idef{org114}Suranaree University of Technology, Nakhon Ratchasima, Thailand
\item \Idef{org115}Technical University of Ko\v{s}ice, Ko\v{s}ice, Slovakia
\item \Idef{org116}Technische Universit\"{a}t M\"{u}nchen, Excellence Cluster 'Universe', Munich, Germany
\item \Idef{org117}The Henryk Niewodniczanski Institute of Nuclear Physics, Polish Academy of Sciences, Cracow, Poland
\item \Idef{org118}The University of Texas at Austin, Austin, Texas, United States
\item \Idef{org119}Universidad Aut\'{o}noma de Sinaloa, Culiac\'{a}n, Mexico
\item \Idef{org120}Universidade de S\~{a}o Paulo (USP), S\~{a}o Paulo, Brazil
\item \Idef{org121}Universidade Estadual de Campinas (UNICAMP), Campinas, Brazil
\item \Idef{org122}Universidade Federal do ABC, Santo Andre, Brazil
\item \Idef{org123}University College of Southeast Norway, Tonsberg, Norway
\item \Idef{org124}University of Cape Town, Cape Town, South Africa
\item \Idef{org125}University of Houston, Houston, Texas, United States
\item \Idef{org126}University of Jyv\"{a}skyl\"{a}, Jyv\"{a}skyl\"{a}, Finland
\item \Idef{org127}University of Liverpool, Liverpool, United Kingdom
\item \Idef{org128}University of Science and Techonology of China, Hefei, China
\item \Idef{org129}University of Tennessee, Knoxville, Tennessee, United States
\item \Idef{org130}University of the Witwatersrand, Johannesburg, South Africa
\item \Idef{org131}University of Tokyo, Tokyo, Japan
\item \Idef{org132}University of Tsukuba, Tsukuba, Japan
\item \Idef{org133}Universit\'{e} Clermont Auvergne, CNRS/IN2P3, LPC, Clermont-Ferrand, France
\item \Idef{org134}Universit\'{e} de Lyon, Universit\'{e} Lyon 1, CNRS/IN2P3, IPN-Lyon, Villeurbanne, Lyon, France
\item \Idef{org135}Universit\'{e} de Strasbourg, CNRS, IPHC UMR 7178, F-67000 Strasbourg, France, Strasbourg, France
\item \Idef{org136} Universit\'{e} Paris-Saclay Centre d¿\'Etudes de Saclay (CEA), IRFU, Department de Physique Nucl\'{e}aire (DPhN), Saclay, France
\item \Idef{org137}Universit\`{a} degli Studi di Foggia, Foggia, Italy
\item \Idef{org138}Universit\`{a} degli Studi di Pavia, Pavia, Italy
\item \Idef{org139}Universit\`{a} di Brescia, Brescia, Italy
\item \Idef{org140}Variable Energy Cyclotron Centre, Homi Bhabha National Institute, Kolkata, India
\item \Idef{org141}Warsaw University of Technology, Warsaw, Poland
\item \Idef{org142}Wayne State University, Detroit, Michigan, United States
\item \Idef{org143}Westf\"{a}lische Wilhelms-Universit\"{a}t M\"{u}nster, Institut f\"{u}r Kernphysik, M\"{u}nster, Germany
\item \Idef{org144}Wigner Research Centre for Physics, Hungarian Academy of Sciences, Budapest, Hungary
\item \Idef{org145}Yale University, New Haven, Connecticut, United States
\item \Idef{org146}Yonsei University, Seoul, Republic of Korea
\end{Authlist}
\endgroup
\end{document}